\documentclass[accepted]{uai2021} % after acceptance, for a revised
\usepackage[american]{babel}
\usepackage{natbib} % has a nice set of citation styles and commands
\usepackage{mathtools} % amsmath with fixes and additions
\usepackage{booktabs} % commands to create good-looking tables
\usepackage{tikz} % nice language for creating drawings and diagrams

\usepackage{balance}

\usepackage{wrapfig}

\usepackage{caption}
\usepackage[separate-uncertainty=true]{siunitx}
\usepackage{mathtools}
\usepackage{graphicx}
\usepackage{courier}
\usepackage{epstopdf}
\usepackage{color}
\usepackage{csquotes}

\usepackage{amsmath}
\usepackage{amsfonts}
\usepackage{amssymb}
\usepackage[subnum]{cases}
\usepackage{booktabs} %For formal tables
\usepackage{pifont}

\usepackage{algorithm,algorithmicx,algpseudocode}

\usepackage{pifont}% http://ctan.org/pkg/pifont

\usepackage {tikz}
\usetikzlibrary {positioning}
\definecolor {processblue}{cmyk}{0.96,0,0,0}
\usepackage{lipsum,adjustbox}

\usepackage[subnum]{cases}
\usepackage{color,soul}

\usepackage{subcaption}

\usepackage{footnote}
\makesavenoteenv{tabular} %ICDE version does not support

\usepackage{multirow}

\usepackage{booktabs,subcaption,amsfonts,dcolumn}
\newcolumntype{d}[1]{D..{#1}}
\usepackage{xcolor}

\usepackage{textcomp}

\title{Unsupervised Constrained Community Detection via \\Self-Expressive Graph Neural Network}

\author[1]{Sambaran Bandyopadhyay \thanks{Sambaran was affiliated to Indian Institute of Science and IBM Research when the work was conducted.}}
\author[2]{Vishal Peter \thanks{Both the authors contributed equally.}}
\affil[1]{%
    Amazon Science\\
    Bangalore, India
}
\affil[2]{%
    Indian Institute of Science\\
    Bangalore, India
}
\begin{document}
\maketitle

\begin{abstract}
  Graph neural networks (GNNs) are able to achieve promising performance on multiple graph downstream tasks such as node classification and link prediction. Comparatively lesser work has been done to design GNNs which can operate directly for community detection on graphs. Traditionally, GNNs are trained on a semi-supervised or self-supervised loss function and then clustering algorithms are applied to detect communities. However, such decoupled approaches are inherently sub-optimal. Designing an unsupervised loss function to train a GNN and extract communities in an integrated manner is a fundamental challenge. To tackle this problem, we combine the principle of self-expressiveness with the framework of self-supervised graph neural network for unsupervised community detection for the first time in literature. Our solution is trained in an end-to-end fashion and achieves state-of-the-art community detection performance on multiple publicly available datasets.
\end{abstract}

\section{Introduction}\label{sec:intro}
Graphs or networks are ubiquitous in our daily life. Graph representation learning \citep{perozzi2014deepwalk,hamilton2017representation} is the task of mapping different components of a graph (such as nodes, edges or the entire graph) to a vector space to facilitate downstream graph mining tasks. Among various types of graph representation techniques, graph neural networks (GNNs) \citep{wu2020comprehensive} have received significant attention as they are able to apply neural networks directly on the graph structure. Most of the GNNs can be represented in the form of a message passing network, where each node updates its vector representation by aggregating messages from neighboring nodes with its own \citep{gilmer2017neural,hamilton2017inductive}. GNNs are traditionally trained in a semi-supervised way \citep{kipf2017semi} on a node classification loss when a subset of node labels are available. More recently, unsupervised and self-supervised graph neural networks have been proposed where a reconstruction loss \citep{kipf2016variational,bandyopadhyay2020outlier} or noise contrastive loss \citep{velivckovic2018deep,zhu2020deep} is used to train the networks.

Community detection is one of the most important tasks for network analysis and has been studied for decades in classical network analysis community \citep{fortunato2016community,xie2013overlapping}. However, compared to other tasks such as node classification \citep{kipf2017semi,velivckovic2018deep} and link prediction \citep{kipf2016variational,zhang2018link}, community detection has not been explored much in the framework of graph neural networks. Being inherently unsupervised in nature, it is challenging to train GNNs for community detection directly. Traditionally, methods have been proposed where a graph representation learning algorithm is trained on a generic unsupervised loss and then a clustering algorithm is applied as a post-processing step to discover communities \citep{perozzi2014deepwalk,bandyopadhyay2019outlier}. Such approaches are sub-optimal in nature as the node representation learning module and the clustering algorithm work independently. More recently, there have been efforts to train graph neural networks directly for community detection in graphs \citep{bo2020structural,zhangcommunity} (Section \ref{sec:related}).

In contrast to a fully unsupervised approach, a graph neural architecture is proposed in \citep{chen2019supervised} for a supervised version of community detection . In classical machine learning, constraint clustering has been shown to be very efficient where must-link or no-link constraints are given as input \citep{wagstaff2001constrained}. But, obtaining direct ground truth community labels or such pair-wise constraints is expensive for real-world networks. In this paper, we aim to derive such constraints in an unsupervised way, by using the principle of self-expressiveness of data \citep{ji2014efficient}. This allows to express each data point by a linear combination of other data points which potentially lie in the same subspace. The principle of self-expressiveness has been successfully applied in computer vision and image processing for object detection and segregation \citep{zhang2019self,li2015structured}. However, inherent computational demand to build pair-wise similarity matrix and subsequent use of spectral clustering makes it infeasible to directly apply the principle of self-expressiveness and subspace clustering in domains like graphs where number of nodes can be very large \citep{elhamifar2009sparse,ji2014efficient,ji2017deep}. We have taken a different approach in this paper to address these computational challenges. Our solution uses a self-supervised GNN and generate node communities from the embeddings obtained. To guide the generated communities, we use the principle of self-expressiveness on randomly sampled batches of nodes to generate a set of soft must-link and no-link constraints.

Following are the novel contributions made in this paper:
\begin{itemize}
    \item We propose a novel community detection algorithm, referred as \textit{SEComm} (\underline{S}elf-\underline{E}xpressive \underline{Comm}unity detection in graph). To the best of our knowledge, we are the first in literature to combine the principle of self-expressiveness with the framework of self-supervised graph neural network for unsupervised community detection. Our solution is able to use both link structure and the node attributes of a graph to detect node communities.
    \item To address the computational issues, our solution uses the principle of self-expressiveness to generate a set of \textit{soft must-link or no-link constraints} on a subset of nodes divided into batches. In contrast to existing literature on self-expressiveness (which typically applies spectral clustering as a post-processing step), our solution is trained in an end-to-end fashion.
    \item To show the merit of the proposed algorithm, we conduct experiments with multiple publicly available graph datasets and compare the results with a diverse set of algorithms. SEComm is able to improve the state-of-the-art performance of unsupervised community detection with a significant margin in almost all the real-world datasets we used. Model ablation study and sensitivity analysis give further insights of the algorithm. Source code of SEComm is available at \url{https://github.com/viz27/SEComm}. 
\end{itemize}

\section{Related Work}\label{sec:related}
We have presented the related work into three categories.

\textbf{Graph Neural Networks}: Graph neural networks have gained significant attention in last few years with their success on a diverse set of applications \citep{wu2020comprehensive,desai2021graph}. Typically, GNNs are trained on node-classification, link prediction and graph reconstruction losses \citep{kipf2017semi,hamilton2017inductive}. Recently, self-supervised learning has been able to achieve performance close to supervised learning for multiple downstream tasks \citep{belghazi2018mutual,hjelm2018learning}. Extending the concept of information maximization, DGI \citep{velivckovic2018deep} and GRACE \citep{zhu2020deep} have been proposed where information between different graph entities (graph-level to node-level, corrupted versions of a graph etc.) are maximized. However, none of the GNNs above handles community detection in their respective objectives.

\textbf{Principle of Self-Expressiveness}: The concept of self expressiveness was proposed to cluster data drawn from multiple low dimensional linear or affine subspaces embedded in a high dimensional space \citep{elhamifar2009sparse}. Given enough samples, each data point in a union of subspaces can always be written as a linear or affine combination of all other points \citep{elhamifar2009sparse,ji2014efficient}. 
%By searching for the sparsest combination for each data point, one can obtain other points lying in the same subspace.
%Different types of matrix norms have been used to recover such sparse representation and equivalence between them under some mild conditions are studied in the literature. \citep{elhamifar2009sparse,ji2014efficient}.
Subspace clustering exploits this to build a similarity matrix, from which the segmentation of the data can be easily obtained using spectral clustering \citep{lu2012robust,ji2014efficient}. 
Recently, a deep learning based subspace clustering method has been proposed where an encoder is used to map data to some embedding space before building the pair-wise similarity matrix and applying spectral clustering \citep{ji2017deep}.
%Principle of self-expressiveness has been used successfully for domains like computer vision and image processing \citep{zhang2019self,li2015structured}. 
However, inherent computational demand to build pair-wise similarity matrix and subsequent use of spectral clustering makes it infeasible to directly apply the principle of self-expressiveness in domains like graphs where the number of nodes can be very large.

\textbf{Community Detection with GNNs}: %Graph modularity \citep{newman2006modularity} and matrix factorization based approaches \citep{yang2013overlapping} have traditionally been used for community detection in graphs. These methods primarily use link structure of the networks to derive different types of similarity metrics between the nodes to detect communities. Unfortunately, they often use shallow architectures, do not use rich node attributes and are not scalable to larger graphs.
% Traditionally, representation learning and graph neural networks can be employed to obtain node embeddings, followed by the application of some clustering algorithm (like k-means++) to detect embeddings . 
As explained in Section \ref{sec:intro}, a disjoint approach of applying clustering on node embeddings obtained by some representation algorithm is inherently sub-optimal in nature \citep{bandyopadhyay2020outlier}.
In \citep{zhang2019attributed}, the authors have used an adaptive graph convolution method that performs high-order graph convolution to obtain smooth node embeddings that capture global cluster structure. The node embeddings obtained are subsequently used to detect communities using spectral clustering.
In \citep{sun2019vgraph}, a probabilistic generative model is proposed to learn community membership and node representation collaboratively.
More recently, researchers have tried to propose GNN algorithms that can operate directly for community detection in a graph \citep{bo2020structural}. 
% Community detection on multi-view graph is proposed in \citep{chengmulti}. 
In \citep{zhangcommunity}, authors propose to derive node community membership in the hidden layer of an encoder and introduced a community-centric dual decoder to reconstruct network structures and node attributes in an unsupervised fashion. Our work is towards this direction of obtaining node communities directly in the framework of graph neural networks.

\section{Problem Formulation}\label{sec:prob}
Let us denote an input graph by $G=(V,E,X)$, where $V = \{1,2,\cdots,N\}$ is the set of nodes and $E \subseteq V \times V$ is the set of edges. We assume that each node has some attribute values present in a vector $x_i \in \mathbb{R}^F$ and $X = [x_1 \; x_2 \; \cdots \; x_N]^T \in \mathbb{R}^{N \times F}$ is the node attribute matrix of the graph. The goal of our work is to learn a function $f: V \mapsto [K]$, where $[K] = \{1,2,\cdots,K\}$ is the set of community (or cluster) indices, to map each node to a community by exploiting the link structure and node attributes of the graph. We want to achieve this without having any ground truth community information of a node. Intuitively, nodes which are closely connected in the graph or have similar attributes, should be members of the same community.
Important notations in the paper are summarized in Table \ref{tab:notations}.
\begin{table}[H]
\centering
\resizebox{\linewidth}{!}{%
\begin{tabular}{*6c}
	\toprule
	\sffamily{Notations} & Explanations\\
%    \sffamily{} & & & & Words & Distribution & links \\
    \hline
	\midrule
	$G = (V,E,X)$ & Input graph \\
	$i,j \in \{1,2,\cdots,N\}=V$ & Indices over nodes \\
	$X \in \mathbb{R}^{N \times F}$ & Node feature matrix \\
	$Z \in \mathbb{R}^{N \times F'}$ & Node embedding matrix \\
	$M,\; P$ & Batch size and number of batches sampled\\
	$S_{ij} \in [0,1]$ & Similarity between nodes $i$ and $j$ \\
	$\mathfrak{S} \subseteq V \times V$ & node pairs whose similarities are computed\\
	$\mathfrak{S}_{ext} \subseteq \mathfrak{S}$ & After filtering out with $\theta_{low}$ and $\theta_{high}$\\
	$C_i \in \mathbb{R}^K$ & Community membership vector for $i$th node \\
	$C \in \mathbb{R}^{N \times K}$ & Output node community membership matrix \\
	$W_{SS}$, $W_{MLP}$ & Parameters of SS-GNN and MLP modules\\
    \bottomrule
	\end{tabular}
	}
\caption{Different notations used in the paper}
\label{tab:notations}
\end{table}

\begin{figure}[h!]
  \centering
  \includegraphics[width=\linewidth]{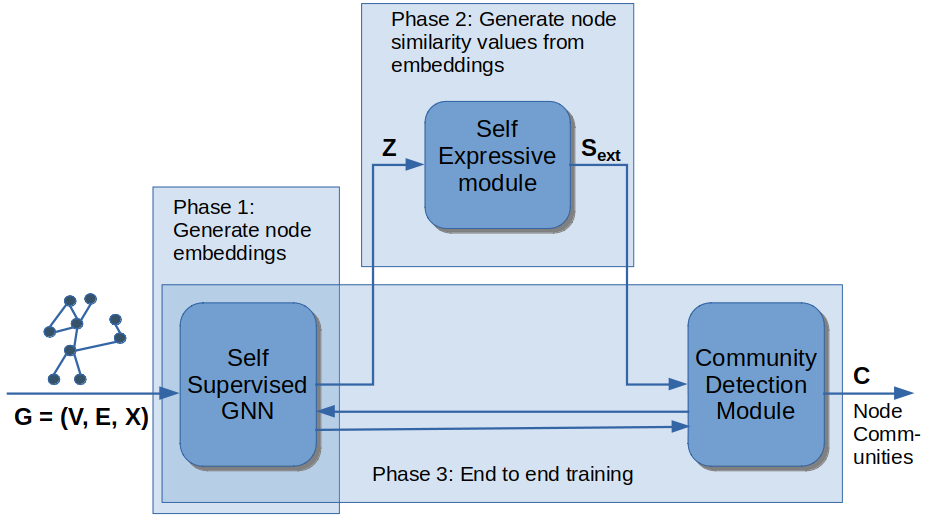}
  \caption{Training phases of SEComm}
  \label{fig:solution}
\end{figure}

\section{Our Solution: SEComm}\label{sec:soln}
There are multiple steps in our proposed solution SEComm as shown in Figure \ref{fig:solution}. We discuss each of them.

\subsection{Self-supervised Node Embedding}\label{sec:nodeEmb}
The first step of SEComm is to learn node representation in an unsupervised way. Self-supervised learning \citep{hjelm2018learning} has been used recently for obtaining both node embeddings \citep{velivckovic2018deep,zhu2020deep} and graph-level embeddings \citep{Sun2020InfoGraph:}. Potentially, any self-supervised differentiable approach to obtain node representation can be integrated with our solution. In the following, we have adopted the principle of mutual information maximization between two corrupted versions of the given graph, motivated from \citep{velivckovic2018deep,zhu2020deep}, which is then used to formulate the final objective of SEComm in Section \ref{sec:NodeComm}.

Given the input graph $G=(V,E,X)$, two graph views $G_1$ and $G_2$ are generated from it by employing a corruption function. The corruption function randomly removes a small portion of edges from the input graph and also randomly masks a fraction of dimensions with zeros in node features. The vertex sets of $G_1$ and $G_2$ remain the same. These views are used for contrastive learning at both graph topology and node feature levels. We use a GCN encoder to generate node embeddings for both $G_1$ and $G_2$. For a graph $G$, GCN encoder derives node representations as follows:
\begin{align}\label{eq:encoder}
    Z = f(X, A) = \text{ReLU}(\hat{A} \; \text{ReLU}(\hat{A}XW^{(0)}) \; W^{(1)})
\end{align}
where each row of $Z \in \mathbb{R}^{|V| \times F'}$ contains the corresponding node representation. $A$ is the adjacency matrix of the graph $G$. We compute $\Tilde{A} = A + I$, where $I \in \mathbb{R}^{|V| \times |V|}$ is the identity matrix and the degree diagonal matrix $\Tilde{D_{ii}}$ with $\Tilde{D_{ii}} = \sum\limits_{j \in V} \Tilde{A}_{ij}$, $\forall i \in V$. We set $\hat{A} = \Tilde{D}^{-\frac{1}{2}} \Tilde{A} \Tilde{D}^{-\frac{1}{2}}$.
%The form of propagation rule has been achieved by a first-order approximation of localized spectral filters on graphs \citep{kipf2017semi,defferrard2016convolutional}.
$W^{(0)}$ and $W^{(1)}$ are the trainable parameter matrices of GCN. Let us use $Z_1$ and $Z_2$ to denote the node embedding matrices for the two views $G_1$ and $G_2$ obtained from the GCN encoder (parameters shared).

Next, the following noise contrastive objective (via a discriminator) is used. For any node $i \in V$, let us denote the corresponding nodes in $G_1$ and $G_2$ as $G_1(i)$ and $G_2(i)$ respectively. For each $i \in V$, the pair $(G_1(i), G_2(i))$ is considered as a positive example. Negative examples are sampled from both the views for each node $i \in V$. More formally, we randomly select a set of nodes $V_{- i} = \{j \in V \;|\; j \neq i\}$ such that $|V_{- i}| = N_{-}$ (number of negative samples), $\forall i \in V$. Both $(G_1(i), G_1(j))$ and $(G_1(i), G_2(j))$ are considered as negative examples. Then, the following objective function is minimized:
\begin{equation}\label{eq:SS}
\begin{aligned}
&\underset{W_{SS}}{\text{min}} \; \mathcal{L}_{SS} \; = \; &&\sum\limits_{i \in V} \Bigg[ - \frac{cos(Z_{1i},Z_{2i})}{\tau} \\
&&& + \log \Big( \sum\limits_{j \in V_{- i}} e^{\frac{cos(Z_{1i},Z_{1j})}{\tau}} + e^{\frac{cos(Z_{1i},Z_{2j})}{\tau}} \Big) \Bigg]
\end{aligned}
\end{equation}
where $Z_{1i}$ and $Z_{2i}$ denotes the $i$th row of $Z_1$ and $Z_2$ respectively, $cos()$ is the cosine similarity between the two embeddings and $\tau$ is a temperature parameter. This essentially maximizes the agreement between the embeddings of $i$th node in two views. Both link structure and node attributes of the graph are considered because of the use of GCN encoder in Equation \ref{eq:encoder}. Node embedding matrix $Z$ of the original input graph $G$ can be obtained by Equation \ref{eq:encoder}, where the purpose of $G_1$ and $G_2$ were just to enable the training of the self-supervised loss in Equation \ref{eq:SS}.

\subsection{Learning Node Similarities through Self-Expressiveness}\label{sec:selfExp}
Vector node representations obtained in Section \ref{sec:nodeEmb} are quite generic. There is no guarantee that similarities between the nodes captured through such embeddings are suitable to discover communities in the graph. In contrast to node classification or other supervised tasks, lack of any ground truth in training for community detection makes the problem highly non-trivial. To tackle this, we use the principle of self-expressiveness \citep{elhamifar2009sparse,ji2017deep} which aims to approximate a data point by a sparse linear sum of a subset of other points, which stays in the same subspace. This is more prevalent when the graph has large number of nodes and embedding dimension is also reasonably high \citep{elhamifar2009sparse}. Based on the contribution of a point to reconstruct some other point, it is possible to learn a pair-wise similarity matrix using this principle. Such a pair-wise similarity can guide the generation of communities from the node embeddings obtained by the self-supervised layer. However, computation of pair-wise similarity matrix for a graph can be too expensive as it needs $O(N^2)$ storage and processing. Hence, we propose a batch-wise learning procedure, as discussed later.

Given the node embedding matrix $Z \in \mathbb{R}^{N \times F'}$ obtained from Section \ref{sec:nodeEmb}, we want to derive a node similarity matrix $S \in \mathbb{R}^{N \times N}$ using the principle of self-expressiveness. For each node $i \in V$, we first try to express $Z_i$ ($i$th row of $Z$) by a linear sum of few other node embeddings $Z_j$, $j \neq i$. So, $Z_i = \sum\limits_{j \in V} q_{ij} Z_j$, where $q_{ij}$ is the $(i,j)$th element of a coefficient matrix $Q \in \mathbb{R}^{N \times N}$ and we enforce $Q_{ii}=0$ to avoid the trivial solution of Q being assigned to a identity matrix. We need to learn this coefficient matrix $Q$ which will be used to generate similarity matrix $S$.
It can be shown \citep{ji2014efficient} under the assumption of subspace independence that, by minimizing certain norms of $Q$, it is possible to have a block-diagonal structure (up to a permutation) of $Q$. In that case, each block in $Q$ would contain nodes which belong to the same subspace. This can be posed as the following optimization problem.
\begin{equation}
\begin{aligned}
\underset{Q}{\text{min}} \; ||Q||_p \;\; \text{such that,} \; Z=QZ; \; \text{diag}(Q)=0
\end{aligned}
\end{equation}
where $||Q||_p$ is $p$th matrix norm of $Q$ and diag$(Q)$ denotes the diagonal entries of $Q$. Based on the choice in some existing literature \citep{lu2012robust,ji2014efficient}, we use square Frobenius matrix norm for our implementation. However, exact reconstruction of $Z$ using the this principle may not be possible. So, we relax the hard constraint $Z=QZ$ with square Frobenius norm of $(Z - QZ)$ (soft constraint). This gives us the following objective function. 
\begin{equation}\label{eq:SE}
\begin{aligned}
\underset{Q}{\text{min}} \; \mathcal{L}_{SE} \; = \; || Z - QZ ||_F^2 + \lambda_1 ||Q||_F^2 \;\; \text{such that,} \; \text{diag}(Q)=0
\end{aligned}
\end{equation}
where $\lambda_1$ is a weight parameter of this optimization.

%The matrix $Q$ learned in this way can be used to construct a symmetric node similarity matrix $S$. 
In principle, while a pairwise similarity matrix $S$ can be constructed trivially as $Q$+$Q^T$, many heuristics have been proposed to improve the clustering performance of $S$ (when using methods such as spectral clustering directly on $S$). We follow the heuristics proposed in \citep{ji2014efficient} to construct the node similarity matrix $S$ as:
\begin{enumerate}
\item $Q' = \frac{1}{2}(Q + Q^T)$
\item Compute the $r$ rank SVD of $Q'$, ie. $Q' = U \Sigma V^T$, where $r = dK+1$, $K$ is the number of communities and $d$ is the maximal intrinsic dimension of subspaces which is set to 4 in all our experiments.
\item Compute $L = U \Sigma^\frac{1}{2}$ and normalize each row of $L$ to have unit norm.
\item Set negative values in $L$ to zero to obtain $L'$.
\item Construct similarity matrix $S$ as $S = (L' + L'^T)/||L||_\infty$, so that $s_{ij} \in [0,1]$.
\end{enumerate}

As mentioned before, an inherent difficulty to compute the pair-wise similarity matrix is the computation and storage of $N \times N$ dimensional matrix $S$. So, instead of computing this matrix for all pairs of nodes, we use batch-wise learning. We sample batches of randomly selected nodes with batch size $M$, where $M \leq N$. We train the loss in Equation \ref{eq:SE} for each batch. The required computation in each batch is $O(M^3)$ (for solving Equation \ref{eq:SE} and the subsequent use of SVD decomposition) which is much lesser than $O(N^3)$ for a significantly smaller $M$. However, the problem with this approach is that one would not get complete similarity matrix for the graph. It only computes $s_{ij}$ if nodes $i$ and $j$ belong to a same batch. Let us denote $\mathfrak{S}$ to be the set of node pairs for which the similarity is computed in the batch-wise learning. Clearly, $\mathfrak{S} \subsetneq V \times V$ and $|\mathfrak{S}| << N^2$ (when $M<N)$. This makes it difficult to use with spectral clustering, as most of the subspace clustering algorithms do \citep{ji2017deep}. But as explained next, our overall solution does not need all the node-pair similarities. Rather, it filters the existing similarities computed with the batch-wise solution using a simple trick explained next.

\subsection{Constrained Node Community Detection}\label{sec:NodeComm}
Instead of applying expensive spectral clustering on the complete matrix $S$ as a post processing step to find node clusters, we use a neural network based solution which is significantly more scalable. We use a fully-connected multilayer perceptron (MLP) with the set of trainable parameters $W_{MLP}$ to map each node embedding to its corresponding soft community membership vector as follows.
\begin{equation}\label{eq:clus}
    C_i = \text{Softmax}(\text{MLP}(Z_i)) \in \mathbb{R}^K
\end{equation}
where the MLP maps each $Z_i \in \mathbb{R}^{F'}$ to a $K$ dimensional vector, $K$ is the number of communities. We assume to know $K$ beforehand. The softmax layer converts the $K$ dimensional vector to a probability distribution such that $c_{ik}$ ($k$th element of $C_i$) denotes the probability that $i$th node belongs to $k$th community, $\forall k = [K]$. Equation \ref{eq:clus} ensures that nodes having similar embeddings will be mapped to similar positions in the $(K-1)$ dimensional probability simplex. However, relying completely on embeddings to detect communities is not desirable since the embeddings are generated with generic objectives. So, they may not be optimal to generate node communities. Hence, we use the information learned in node-pair similarities in Section \ref{sec:selfExp} to guide both the detection of node communities by training the parameters of MLP in Equation \ref{eq:clus} and updating node embeddings.

Let us form the community membership matrix $C = [C_1,\cdots,C_N]^T$ $\in \mathbb{R}^{N \times K}$. If the complete node similarity matrix $S$ is available, one may try to minimize the following objective.
\begin{equation}\label{eq:clusloss1}
\begin{aligned}
\underset{W_{SS},\; W_{MLP}}{\text{min}} || C C^T - S ||_F^2
\end{aligned}
\end{equation}
There are multiple drawbacks present in the objective function above. First, it needs us to compute the complete node similarity matrix $S$ in Section \ref{sec:selfExp} which prevents the batch-learning mechanism explained before. Next, the computation involved is $O(N^2)$ in Equation \ref{eq:clusloss1}. Further, there is another issue if one wants to use all pair-wise node similarities in $S$ to guide the community detection. Due to noise in the dataset, many of the pair-wise similarities may not reflect the actual similarities between the nodes. The similarity values which are around 0.5 neither express a strong similarity nor a strong dissimilarity between a node pair. So they are less informative compared to the similarity values which are close to 0 or 1. But they can still influence the parameters of the neural network because of Equation \ref{eq:SE}.

Hence, instead of considering all the pair-wise similarity values, we only consider the ones in $\mathfrak{S}$ computed over the batches as discussed in Section \ref{sec:selfExp}. Further, we have observed experimentally in Section \ref{sec:sensitivity} that for a larger dataset, it is okay even if some nodes are not part of any of the batches selected randomly. Thus, the number of batches can be significantly smaller than $\frac{N}{M}$ for a larger dataset.
%In our experiments, we keep number of batches to be $0.5 \times \frac{N}{M}$ for each dataset.
We also introduce two thresholds $\theta_{low}$ and $\theta_{high}$ to use only those node-pair similarities which are extreme in their values, thus more informative in nature. We set $0 < \theta_{low} \leq \theta_{high} < 1$. We also set $\theta_{high} = 1 - \theta_{low}$, as this choice works well in the experiments and also reduces the number of hyperparameters.
Let us introduce the set $\mathfrak{S}_{ext} \subseteq \mathfrak{S}$ as follows.
\begin{equation}\label{eq:S_ext}
    \mathfrak{S}_{ext} = \Big\{ (i,j) \in \mathfrak{S} \;|\; S_{ij} \leq \theta_{low} \;\text{or}\; S_{ij} \geq \theta_{high} \Big\}
\end{equation}
Here, a node pair $(i,j)$ in $\mathfrak{S}_{ext}$ should be roughly constrained to be in the same cluster when $S_{ij}$ value is very high or in different clusters when $S_{ij}$ is very low. Thus, we derive a set of soft version of must-link and no-link constraints in an unsupervised way to guide the formation of communities.
%This type of constraints have been useful to boost the performance for constraint clustering scenario \citep{wagstaff2001constrained,basu2004active}. But unlike them, we are learning such (soft) constraints using the principle of self-expressiveness in a scalable fashion. 
With these, we formulate the following optimization to detect the communities:
\begin{equation}\label{eq:clusloss2}
\begin{aligned}
\underset{W_{SS},\; W_{MLP}}{\text{min}} \sum\limits_{(i,j) \in \mathfrak{S}_{ext}} \Big( C_i^T C_j - S_{ij} \Big)^2
\end{aligned}
\end{equation}
%This would ensure that two nodes with a high similarity value $S_{ij}$ should have similar community memberships and vice-versa. 
By considering only the node pairs in $\mathfrak{S}_{ext}$, we are able to ignore the pairs which are neither too similar nor too dissimilar, to contribute to the learning of community memberships.
As $C_i$ is a probability distribution over all the $K$ communities for a node $i$, we want to avoid trivial community formations where each node is assigned to all the communities with roughly uniform probabilities, or all the nodes are assigned to a single community \citep{bianchi2020spectral}. So, we update the main objective in Equation \ref{eq:clusloss2} as:
\begin{equation}\label{eq:cluslossFinal}
\begin{aligned}
&\underset{W_{SS},\; W_{MLP}}{\text{min}} \; \mathcal{L}_{Com} = &\sum\limits_{(i,j) \in \mathfrak{S}_{ext}} \Big( C_i^T C_j - S_{ij} \Big)^2 \\
&&+ \lambda_2 \biggl\lvert \biggl\lvert \frac{C^T C}{||C^T C||_F} - \frac{I_K}{\sqrt{K}} \biggl\lvert \biggl\lvert_F^2
\end{aligned}
\end{equation}
The second component in the equation above ensures that communities are close to orthogonal and they are balanced in size. Please note that due to the use of neural network to generate community membership for each node in Equation \ref{eq:clus}, the optimization in Equation \ref{eq:cluslossFinal} is not a discrete optimization. Rather, we solve it with respect to the parameters $W_{SS}$ of the self-supervised layer (Eq. \ref{eq:SS}) and $W_{MLP}$ of the MLP (Eq. \ref{eq:clus}). 
%Thus, we are able to solve the unsupervised community detection in the framework of neural network. 
The total loss to train the node embeddings and community detection can be written as a weighted sum of self-supervised loss and community detection loss, which is shown below:
\begin{equation}\label{eq:SECommloss}
\begin{aligned}
\underset{W_{SS},\; W_{MLP}}{\text{min}} \; \mathcal{L}_{total} = \alpha\mathcal{L}_{SS} +  \mathcal{L}_{Com}
\end{aligned}
\end{equation}
where $\alpha$ is a weight factor of the optimization. The node-pair similarity values are obtained by solving the batch-learning technique in Section \ref{sec:selfExp}. The overall algorithm SEComm proceeds in an iterative way by solving the self-expressive layer for each batch and then updating the parameters of the neural network by minimizing Equation \ref{eq:SECommloss}. The pseudo code of SEComm is presented in \ref{alg:SEComm}.

\begin{algorithm} %[H]%[tb]
%\thisfloatpagestyle{empty}
%\algsetup{linenosize=\tiny}
  \small
  \caption{\textbf{SEComm} - Self-Expressive Community Detection}
  %\resizebox{0.47\textwidth}{!} {
  \label{alg:SEComm}
%\resizebox{0.5\textwidth}{!}{\begin{minipage}{\textwidth}   
\begin{algorithmic}[1]
      
	\Statex \textbf{Input}: The graph $G=(V,E,X)$, $|V|=N$, $K$: Number of communities in the graph, $M$: Batch size for the self-expressive layer, $P$: Number of batches used for training self expressive layer, Thresholds $\theta_{low}$ and $\theta_{high}$.
    \Statex \textbf{Output}: Community membership vector $C_i \in \mathbb{R}^{K}$ for each node $i \in V$. 
	\State Initialize the parameters of the self-supervised GNN and clustering MLP (in Eq. \ref{eq:clus}).
	\State pre-training step: Obtain node embeddings $Z \in \mathbb{R}^{N \times F'}$ by training the self-supervised GNN.
	\State Initialize $\mathfrak{S}$ as empty.
    \For{$batch \in \{1,2,\cdots,P\}$}
        \State Sample a batch of $M$ nodes from $V$
        \State Learn the pair-wise node similarity matrix $S_m$ for the selected nodes by optimizing Eq. \ref{eq:SE}.
        \State Add all the node-pairs from the batch to $\mathfrak{S}$.
    \EndFor
    \State Construct $\mathfrak{S}_{ext}$ according to Eq. \ref{eq:S_ext}
    \For{$iter \in \{1,2,\cdots,T\}$}
	    \State Generate node embedding matrix $Z \in \mathbb{R}^{N \times F'}$ using the self-supervised GNN.
	    \State Generate cluster membership vector $C_i \in \mathbb{R}^{K}$ for each node $i \in V$.
	    \State Update the parameters of the GNN and clustering MLP by optimizing Eq. \ref{eq:SECommloss}
	\EndFor
	\end{algorithmic}
	%}
  \end{algorithm} %sam modi
  
\subsection{Training and Analysis of SEComm}\label{sec:analysis}
We use ADAM with default parameterization to solve the optimization formulations in Equations \ref{eq:SE} and \ref{eq:SECommloss}. For the self-expressive loss in Equation \ref{eq:SE}, we train until the loss saturates. For the total loss in Equation \ref{eq:SECommloss}, we particularly focus on the saturation of the regularization $\biggl\lvert \biggl\lvert \frac{C^T C}{||C^T C||_F} - \frac{I_K}{\sqrt{K}} \biggl\lvert \biggl\lvert_F^2$. Experimentally, using the convergence of this component explicitly as a stopping criteria for SEComm gives slightly better result for all the datasets, than checking the total convergence. But as explained in Section \ref{sec:expSetup}, different components of the loss function have similar contributions. Hence they saturate almost in the same time for most of the cases. This can also be observed in Section \ref{sec:lossAcc}.

\textbf{Time Complexity}: Time complexity of the self-supervised GNN in Section \ref{sec:nodeEmb} is $O(|E| + NFF'N_-)$, where $N_-$ is the number of negative samples used. The self-expressive layer takes another $O(P M^3)$ time, where $P$ is the number of batches sampled, and $M$ is the size of each batch. Finally, generating community membership takes $O(NK^2)$ time because of solving the loss in Equation \ref{eq:cluslossFinal}. Thus, the overall run time of each iteration of SEComm is linearly dependent on the number of nodes and number of edges in the graph.

\section{Experimental Evaluation}\label{sec:exp}
This section presents the details of the experiments that we conducted and the analysis of the results.

\begin{table}
	\centering
    \resizebox{0.9\linewidth}{!}{%
	\begin{tabular}{*5c}
	\toprule
	\sffamily{Dataset} & \#Nodes & \#Edges & \#Features &\#Labels \\
%    \sffamily{} & & & & Words & Distribution & links \\
    \hline
% 	\sffamily{Synthetic} & 600 & 14202 & 3 & None \\
	\sffamily{Cora} & 2,708 & 5,429 & 1,433 & 7 \\
	\sffamily{Citeseer} & 3,327 & 4,732 & 3,703 & 6 \\
	\sffamily{Pubmed} & 19,717 & 44,338 & 500 & 3 \\
	\sffamily{Wiki} & 2,405 & 17,981 & 4,973 & 17 \\
	\sffamily{Physics} & 34,493 & 247,962 & 8,415 & 5 \\
\bottomrule
	\end{tabular}
    }
    \caption{Summary of the datasets used}
	\label{tab:data}
\end{table}

\subsection{Datasets Used}
To show the merit of SEComm, we conduct experiments on 5 publicly available graph datasets \citep{kipf2017semi,zhang2019attributed}. Different statistics of the datasets are summarized in Table \ref{tab:data}. Cora, Citeseer and Pubmed are citation datasets where nodes correspond to papers and are connected by an edge if one cites the other. Wiki is a collection of webpages where nodes are webpages and are connected if one links to other. Physics is a co-authorship network where nodes are authors, that are connected by an edge if they have co-authored a paper \citep{shchur2018pitfalls}.
Each of these datasets have attribute vector associated with each node. They also have ground truth community membership of each node, which we use to evaluate the performance of our proposed and baseline algorithms.

\begin{table*}
\centering
\resizebox{\linewidth}{!}{
\begin{tabular}{c|c|ccc|ccc|ccc|ccc}
    \toprule
    
     Methods & Input & \multicolumn{3}{c}{Cora} &\multicolumn{3}{c}{Citeseer} & \multicolumn{3}{c}{Pubmed} & \multicolumn{3}{c}{Wiki}\\
    \cmidrule{3-14}
     & & Acc\%  & NMI\%  & F1\%   & Acc\%  & NMI\%  & F1\%   & Acc\%  & NMI\%  & F1\%  & Acc\%  & NMI\%  & F1\%    \\
    %\cline{1-10}
    \midrule
    k-means & Feature & 34.65 & 16.73 & 25.42 & 38.49 & 17.02 & 30.47 & 57.32 & 29.12 & 57.35 & 33.37 & 30.20 & 24.51 \\
    Spectral-f & Feature & 36.26 & 15.09 & 25.64 & 46.23 & 21.19 & 33.70 & 59.91 & 32.55 & 58.61 & 41.28 & 43.99 & 25.20 \\
    \midrule
    Spectral-g & Graph & 34.19 & 19.49 & 30.17 & 25.91 & 11.84 & 29.48 & 39.74 & 3.46 & 51.97 & 23.58 & 19.28 & 17.21 \\
    DeepWalk & Graph & 46.74 & 31.75 & 38.06 & 36.15 & 9.66 & 26.70 & 61.86 & 16.71 & 47.06 & 38.46 & 32.38 & 25.74 \\
    DNGR & Graph & 49.24 & 37.29 & 37.29 & 32.59 & 18.02 & 44.19 & 45.35 & 15.38 & 17.90 & 37.58 & 35.85 & 25.38 \\
    \midrule
    GAE & Both & 53.25 & 40.69 & 41.97 & 41.26 & 18.34 & 29.13 & 64.08 & 22.97 & 49.26 & 17.33 & 11.93 & 15.35 \\
    VGAE & Both & 55.95 & 38.45 & 41.50 & 44.38 & 22.71 & 31.88 & 65.48 & 25.09 & 50.95 & 28.67 & 30.28 & 20.49 \\
    MGAE & Both & 63.43 & 45.57 & 38.01 & 63.56 & 39.75 & 39.49 & 43.88 & 8.16  & 41.98 & 50.14 & 47.97 & 39.20 \\
    ARGE & Both & 64.00 & 44.90 & 61.90 & 57.30 & 35.00 & 54.60 & 59.12 & 23.17 & 58.41 & 41.40 & 39.50 & 38.27 \\
    ARVGE & Both & 63.80 & 45.00 & 62.70 & 54.40 & 26.10 & 52.90 & 58.22 & 20.62 & 23.04 & 41.55 & 40.01 & 37.80 \\
    % DONE & & & & & & & & & & & & & \\
    % AdONE & & & & & & & & & & & & & \\
    AGC & Both & 68.92 & 53.68 & 65.61 & 67.00 & 41.13 & \textbf{62.48} & 69.78 & 31.59 & 68.72 & 47.65 & 45.28 & 40.36 \\
    GUCD & Both & 50.5 & 32.3 & NA & 54.47 & 27.43 & NA & 63.13 & 26.98 & NA & NA & NA & NA \\
    \midrule
    SEComm & Both & \textbf{75.92} & \textbf{56.04} & \textbf{73.94} & \textbf{69.82} & \textbf{42.53} & 60.25 & \textbf{74.49} & \textbf{36.50} & \textbf{73.50} & \textbf{53.10} & \textbf{51.38} & \textbf{44.48} \\
    % SEComm & Both & \textbf{75.8}$\pm$0.1 & \textbf{55.9}$\pm$0.1 & \textbf{73.8}$\pm$0.1 & \textbf{69.8}$\pm$0.1 & \textbf{42.5}$\pm$0.1 & 60.3$\pm$0.0 & \textbf{73.3}$\pm$0.3 & \textbf{34.8}$\pm$0.4 & \textbf{72.0}$\pm$0.3 & \textbf{53.4}$\pm$0.3 & \textbf{50.5}$\pm$1.3 & \textbf{45.5}$\pm$0.4 \\
    Rank  & (SEComm) & 1 & 1 & 1 & 1 & 1 & 2 & 1 & 1 & 1 & 1 & 1 & 1 \\
    \bottomrule
\end{tabular}
}
\caption{Performance of Community Detection by SEComm and other baseline algorithms}
\label{tab:commResults}
\end{table*}

\begin{table}
	\centering
    \resizebox{0.85\linewidth}{!}{%
	\begin{tabular}{*5c}
	\toprule
	\sffamily{Methods} & Acc\% & NMI\% & F1\% & Runtime (sec.) \\
%    \sffamily{} & & & & Words & Distribution & links \\
    \hline
% 	\sffamily{Synthetic} & 600 & 14202 & 3 & None \\
	\sffamily{k-means} & 44.20 & 44.46 & 37.63 & 28 \\
	\sffamily{ARGE} & 60.67 & 51.77 & 62.32 & 2112 \\
	\sffamily{ARVGE} & 61.28 & 53.49 & 65.47 & 2221 \\
	\sffamily{AGC} & 75.36 & \textbf{58.19} & 60.72 & 6931 \\
	\sffamily{SEComm} & \textbf{77.93} & 56.08 & \textbf{76.42} & 624 \\
\bottomrule
	\end{tabular}
    }
    \caption{Community Detection on Physics Dataset}
	\label{tab:commPhysics}
\end{table}

\begin{figure}[h!]
  \centering
  \begin{subfigure}[b]{0.49\linewidth}
    \includegraphics[width=\linewidth]{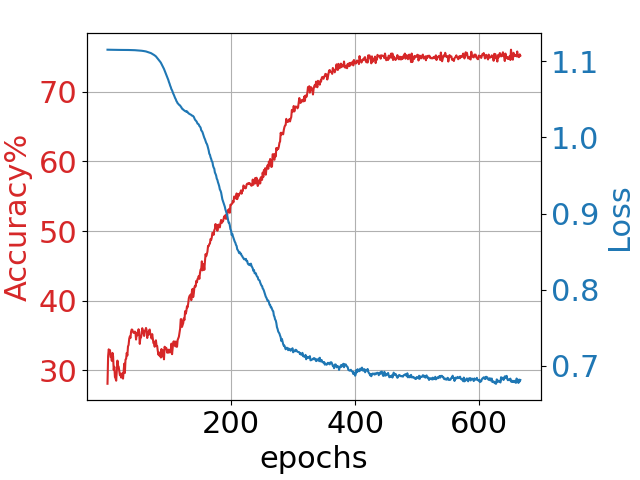}
    \caption{Cora}
    \label{fig:Cora_AccVsLoss}
  \end{subfigure}
  \begin{subfigure}[b]{0.49\linewidth}
    \includegraphics[width=\linewidth]{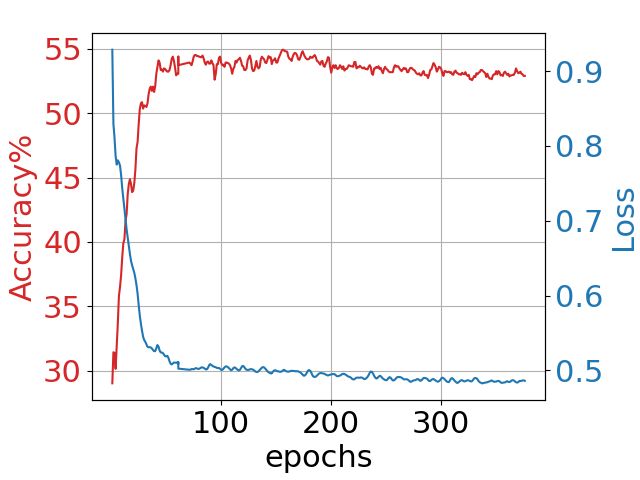}
    \caption{Wiki}
    \label{fig:Wiki_AccVsLoss}
  \end{subfigure}
  \caption{Loss vs Accuracy progression during training}
  \label{fig:AccVsLoss}
\end{figure}

\subsection{Baseline Algorithms}\label{sec:baselines}
We use a diverse set of baselines to compare the performance of SEComm. We divide them into the following categories.

\textbf{Using only Node Features}: As each node is associated with some attribute vectors, we use \textbf{k-means} and spectral clustering (\textbf{Spectral-f}) algorithms directly on the node attributes to cluster nodes into different communities. Naturally, these approaches ignore the graph structure completely.

\textbf{Using only Graph Structure}: We also use spectral clustering (\textbf{Spectral-g}) on the graph structure. Here we consider adjacency matrix of a graph as the similarity matrix between the nodes. We also use popular unsupervised node embedding techniques \textbf{DeepWalk} \citep{perozzi2014deepwalk}, which is a random walk based technique and \textbf{DNGR} \citep{cao2016deep}, which is an auto-encoder based technique.

\textbf{Using both Node Features and Graph Structure}: We use a set of unsupervised graph neural network based techniques. GNN based approaches are naturally able to use both link structure and node attribute of the graph. They are: graph autoencoder (\textbf{GAE}) and graph variational autoencoder (\textbf{VGAE}) \citep{kipf2016variational}, marginalized graph autoencoder (\textbf{MGAE}) \citep{wang2017mgae}, adversarially regularized graph autoencoder (\textbf{ARGE}) and variational graph autoencoder (\textbf{ARVGE}) \citep{pan2018adversarially}.
%Besides, we also use two recently proposed approaches \textbf{DONE} and \textbf{AdONE} \citep{bandyopadhyay2020outlier} which minimize the effect of outlier nodes in graph representation learning.
These methods typically learn the node embeddings and use clustering on the embeddings as a post processing step. Finally, we use two recently proposed community detection methods - \textbf{AGC} \citep{zhang2019attributed}, which uses high-order graph convolution to get node embeddings and detect communities via spectral clustering on the embeddings and \textbf{GUCD} \citep{zhangcommunity}, which uses an auto-encoder based framework to obtain direct community assignments for every node.

\subsection{Experimental Setup}\label{sec:expSetup}
Our proposed algorithm SEComm generates community membership of each node in a graph in the framework of graph neural networks. As each node has a single ground truth community membership in all the datasets that we use, we consider the index of the maximum value of $C_i \in \mathbb{R}^K$ (from Equation \ref{eq:clus}) as the community index of the node generated by SEComm.

There are multiple hyperparameters present in SEcomm. For weight factors in optimization such as $\lambda_1$, $\lambda_2$ and $\alpha$, we check the contribution of different components in a loss function at the beginning of the algorithm, and set these parameters to values such that effective contributions of those components become roughly the same. This ensures that the optimization pays similar importance to different components of SEComm. For the temperature parameter $\tau$, we use the same values used in the literature \citep{zhu2020deep}. 
For threshold parameters $\theta_{low}$ ($0 < \theta_{low} \leq 0.5$), we set it to 0.5 for relatively smaller datasets as we do not want to discard any information for them. For Pubmed, we set it to 0.05 as considering more node-pair similarity values adds noise and also increases runtime of SEComm. However on Physics, the training convergence is not smooth when we set $\theta_{low}$ to a very small number. So, we set it to $0.3$ on this dataset.
As mentioned in Section \ref{sec:NodeComm}, we set $\theta_{high}=1-\theta_{low}$. 
%A summary of all the hyperparameter values are presented in Table \ref{tab:Hyperarams} to ease the reproducibility of the results. 
We have also conducted sensitivity analysis of SEComm with respect to some of these hyperparameters in Section \ref{sec:sensitivity}.

\subsection{Results of Community Detection}
Tables \ref{tab:commResults} and \ref{tab:commPhysics} show the performance of community detection by different baseline algorithms and SEComm. We use three popularly used metrics to evaluate the quality of community detection. They are clustering accuracy (Acc), normalized mutual information (NMI), and macro F1-score (F1) \citep{aggarwal2014data,zhang2019attributed}. We use ground truth community information of the nodes only to calculate these quality metrics. 

While reporting the performance of baseline algorithms for the first four datasets in Table \ref{tab:commResults}, we have collected the best results from the available literature \citep{zhang2019attributed,zhangcommunity} which adopted the same experimental setup. We mark some entry as `NA' if the result of that algorithm for a dataset is not publicly available. For Physics dataset, the baseline results are not available in the literature. So, we have run and reported results only for better-performing and diverse subset of baselines in Table \ref{tab:commPhysics} with adequate hyperparameter tuning. Additionally, we have also reported the runtime on Physics dataset for these algorithms to give more insight about scalability.

We run SEComm 10 times on each dataset and report the average performance. Tables \ref{tab:commResults} and \ref{tab:commPhysics} show that SEComm is able to achieve state-of-the-art (SOTA) performance for all the datasets, and for all the metrics, except on Citeseer-F1\% and Physics-NMI\% scores, where SEComm is next to AGC.
In terms of performance improvement by clustering accuracy, SEComm is able to improve SOTA by 10.1\% on Cora, 4.2\% on Citeseer, 6.7\% on Pubmed, 5.9\% on Wiki and 3.4\% on Physics.
We also check the standard deviation of the performance of SEComm over 10 runs in each dataset. Standard deviation lies in the range of 0.5\% - 1\% on all the datasets, which shows the robustness of SEComm.
Among the baselines, AGC mostly performs better than others. But, AGC is computationally much expensive because of their adaptive strategy for determining k in k-order graph convolution. As shown in Table \ref{tab:commPhysics}, runtime of AGC on Physics dataset is \textasciitilde 16 times more than that of SEComm.
As expected, the algorithms which use both graph structure and node attributes perform better than the ones which use only one of those.
%The better performing baseline AGC does not obtain direct community assignments and uses spectral clustering to detect communities. Whereas, 
The consistent performance of SEComm on all the datasets shows the importance of integrating the objective of community detection directly into the framework of self-supervised graph neural network (the loss from these components propagates to each other through backpropagation). Further, use of the principle of self-expressiveness regularizes the communities formed in SEComm to achieve better performance. Run-time of SEComm and its various components on all the datasets is shown in Table \ref{tab:Runtime}. Usefulness of the individual components of SEComm are presented in Section \ref{sec:ablation}.

\begin{table}[H]
	\centering
    \resizebox{\columnwidth}{!}{%
	\begin{tabular}{*5c}
	\toprule
	\sffamily{Dataset} & GNN & Self-Express. & Comm. Module & Total \\
    \hline
	Wiki & 5.4s & 34.8s & 44.8s & 85.1s \\
    Cora & 6.6s & 39.2s & 84.2s & 130.1s \\
    Citeseer & 27.8s & 58.4s & 157.8s & 244.1s \\
    Pubmed & 239.6s & 103.6s & 121.9s & 465.1s \\
    Physics & 106.9s & 280.4s & 236.4s & 623.7s \\
\bottomrule
	\end{tabular}
    }
    \caption{Runtime of SEComm on various datasets}
	\label{tab:Runtime}
\end{table}

\subsection{Loss and Accuracy of SEComm}\label{sec:lossAcc}
Typically for an unsupervised algorithm, the loss that it minimizes and the metric that is used to evaluate the performance of the algorithm are not necessarily the same. So, it is important to see if reducing the loss over the epochs actually increases the performance of the algorithm with respect to the quality metric. For SEComm, we plot the loss in Equation \ref{eq:SECommloss} and the clustering accuracy that it achieves over different epochs of the algorithm for the datasets Cora and Wiki in Figure \ref{fig:AccVsLoss}. One can see that with the decreasing loss, overall clustering accuracy improves with some minor fluctuations. Thus, the unsupervised loss that SEComm minimizes essentially helps to improve the performance of clustering. This is also another reason of consistent performance (improved metric scores with less standard deviation) of SEComm on multiple datasets.

\begin{figure}[h!]
  \centering
  \begin{subfigure}[b]{0.49\linewidth}
    \includegraphics[width=\linewidth]{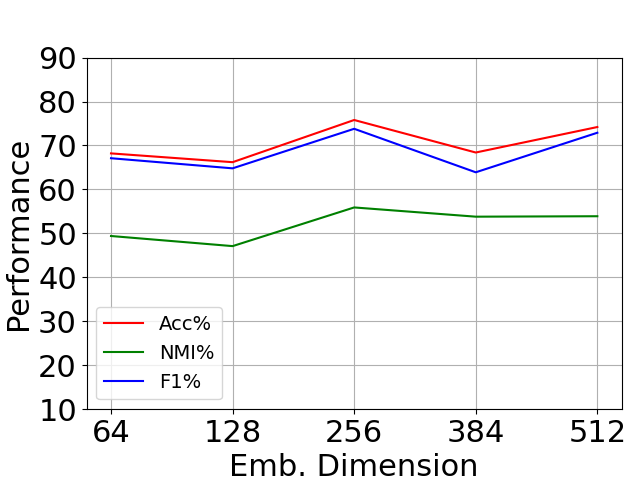}
    \caption{Cora}
    \label{fig:Cora_Embedding}
  \end{subfigure}
  \begin{subfigure}[b]{0.49\linewidth}
    \includegraphics[width=\linewidth]{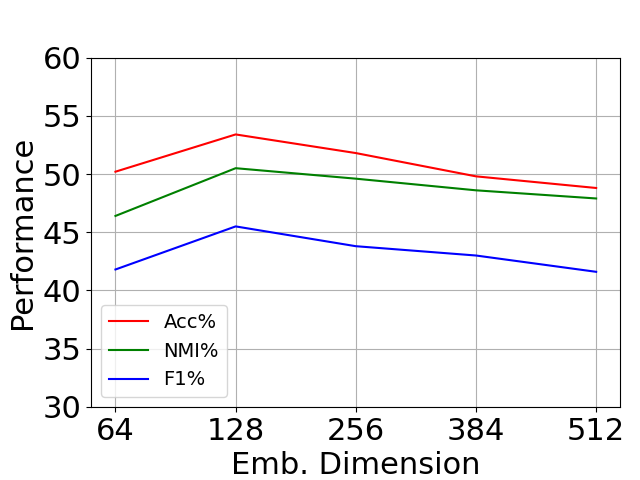}
    \caption{Wiki}
    \label{fig:Wiki_Embedding}
  \end{subfigure}
  \caption{Performance of SEComm with varying embedding dimension}
  \label{fig:embDim}
\end{figure}

\begin{figure}[h!]
  \centering
  \begin{subfigure}[b]{0.49\linewidth}
    \includegraphics[width=\linewidth]{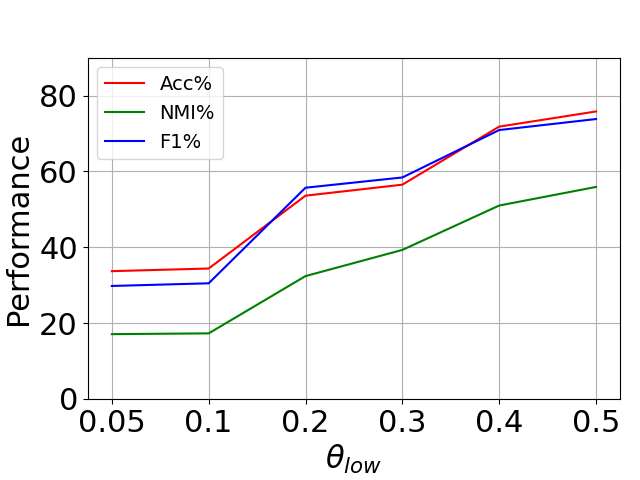}
    \caption{Cora}
    \label{fig:Cora_Threshold}
  \end{subfigure}
%   \begin{subfigure}[b]{0.3\linewidth}
%     \includegraphics[width=\linewidth]{Images/Wiki_Threshold.png}
%     \caption{Wiki}
%     \label{fig:Wiki_Threshold}
%   \end{subfigure}
  \begin{subfigure}[b]{0.49\linewidth}
    \includegraphics[width=\linewidth]{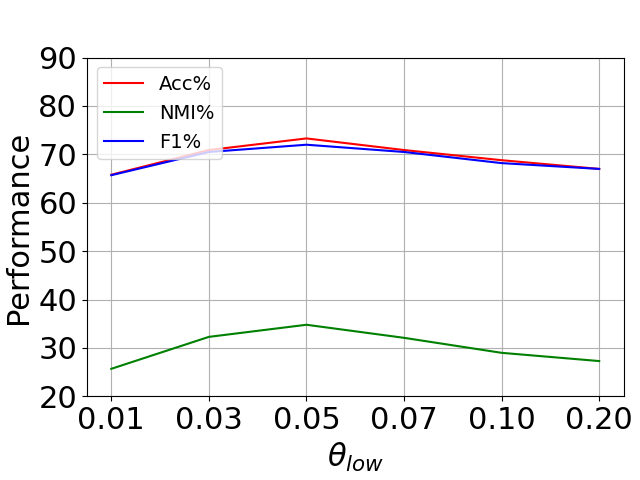}
    \caption{Pubmed}
    \label{fig:PubMed_Threshold}
  \end{subfigure}
  \caption{Performance of SEComm with varying threshold $\theta_{low}$}
  \label{fig:ThresholdVsPerf}
\end{figure}

\begin{figure}[h!]
  \centering
  \begin{subfigure}[b]{0.49\linewidth}
    \includegraphics[width=\linewidth]{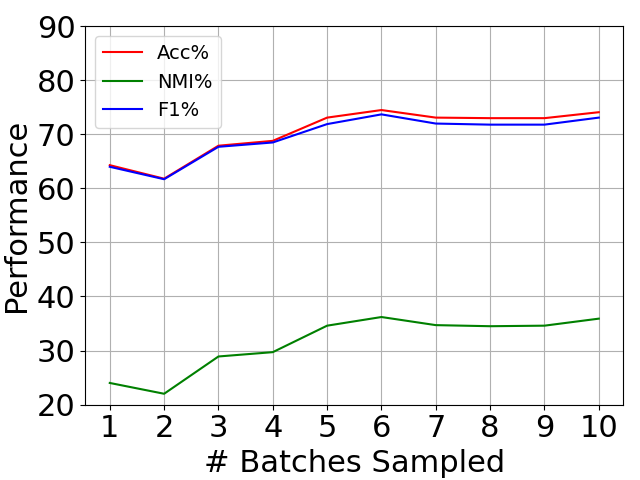}
    \caption{Pubmed}
    \label{fig:batchesSampled}
  \end{subfigure}
  \begin{subfigure}[b]{0.49\linewidth}
    \includegraphics[width=\linewidth]{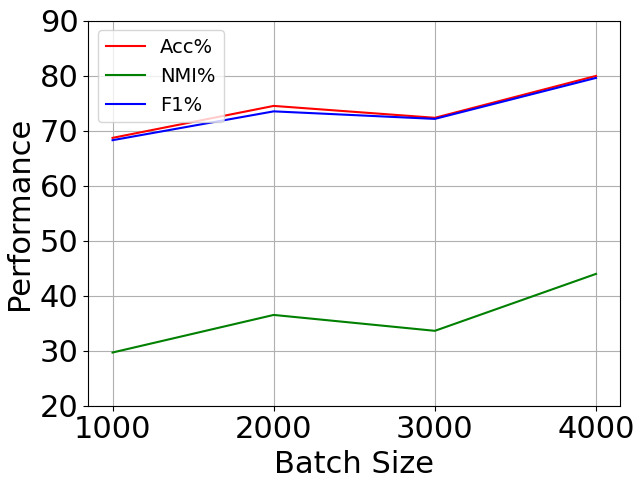}
    \caption{Pubmed}
    \label{fig:batchSize}
  \end{subfigure}
  \caption{Performance of SEComm with varying number of batches sampled and batch size respectively}
  \label{fig:batch}
\end{figure}

% \begin{figure}[h!]
%   \centering
%     \includegraphics[width=0.6\linewidth]{Images/PubMed_DataVariation.png}
%   \caption{Performance of SEComm with varying number of batches sampled in Pubmed}
%   \label{fig:batchPer}
% \end{figure}

\begin{table*}[ht]
\centering
\resizebox{0.9\linewidth}{!}{
\begin{tabular}{c|ccc|ccc|ccc|ccc}
    \toprule
     Methods & \multicolumn{3}{c}{Cora} &\multicolumn{3}{c}{Citeseer} & \multicolumn{3}{c}{Pubmed} & \multicolumn{3}{c}{Wiki}\\
    \cmidrule{2-13}
     & Acc\%  & NMI\%  & F1\%   & Acc\%  & NMI\%  & F1\%   & Acc\%  & NMI\%  & F1\%  & Acc\%  & NMI\%  & F1\%  \\
    \midrule
    SEComm-GNN & 72.85 & 54.40 & 68.16 & 64.95 & 39.01 & \textbf{60.51} & 49.63 & 19.59 & 40.04 & 32.14 & 34.88 & 28.34 \\
    SEComm-Spectral & 74.26 & 55.38 & 71.75 & 57.58 & 36.63 & 54.99 & - & - & - & 49.81 & 51.07 & 43.95 \\
    SEComm-Embeddings & 73.85 & 57.57 & 65.07 & 60.29 & 35.46 & 54.46 & 68.30 & 34.47 & 67.97 & 35.67 & 36.12 & 29.49 \\
    SEComm & \textbf{75.92} & \textbf{56.04} & \textbf{73.94} & \textbf{69.82} & \textbf{42.53} & 60.25 & \textbf{74.49} & \textbf{36.50} & \textbf{73.50} & \textbf{53.10} & \textbf{51.38} & \textbf{44.48} \\
    % SEComm & Both & \textbf{75.8}$\pm$0.1 & \textbf{55.9}$\pm$0.1 & \textbf{73.8}$\pm$0.1 & \textbf{69.8}$\pm$0.1 & \textbf{42.5}$\pm$0.1 & 60.3$\pm$0.0 & \textbf{73.3}$\pm$0.3 & \textbf{34.8}$\pm$0.4 & \textbf{72.0}$\pm$0.3 & \textbf{53.4}$\pm$0.3 & \textbf{50.5}$\pm$1.3 & \textbf{45.5}$\pm$0.4 \\
\bottomrule
\end{tabular}
}
\caption{Model Ablation Study of SEComm}
\label{tab:abalResults}
\end{table*}

\subsection{Sensitivity to Hyperparameters}\label{sec:sensitivity}
In this section, we show the sensitivity of SEComm to different hyperparameters. We keep all other hyperparameters fixed while changing only the hyperparameter of interest.
We vary the embedding dimension $F'$ from 64 to 512 for Cora and Wiki and show the performance of community detection in Figure \ref{fig:embDim}. We can see some fluctuation in the performance for Cora. As other hyperparameters are tuned keeping $F'=256$ for Cora, there are some sudden lows around it. For Wiki, the performance was low at $F'=64$ as that is not sufficient enough to preserve all the information about the graph in the embeddings. It increases at $F'=128$. There is a gradual decrease of performance beyond that as the embeddings start holding noisy information when dimension increases more.

We check the performance on Cora, Wiki and Pubmed in Figure \ref{fig:ThresholdVsPerf} with varying $\theta_{low}$ (we set $\theta_{high} = 1 - \theta_{low}$). As we decrease $\theta_{low}$, we are filtering out more pair-wise similarities, especially the ones which lies in the mid zone of the range $[0,1]$. Filtering out such similarity values might lead to less amount of data to regularize the communities learned by SEComm in Equation \ref{eq:cluslossFinal} for a smaller dataset. Thus, the performance on Cora is affected when $\theta_{low}$ is very low in Figure \ref{fig:ThresholdVsPerf}. But on a larger dataset, filtering out such less informative similarity values (as explained in Section \ref{sec:selfExp}) can lead to removing noise and help improving the performance. Thus in Pubmed, better performance is observed around $\theta_{low} = 0.05$ (which implies $\theta_{high} = 0.95$). Below which the amount data becomes too less to train the algorithm properly, and above which it was adding noise.

As discussed in Section \ref{sec:selfExp}, we do not need to train the self-expressive layer (in Eq. \ref{eq:SE}) on the complete dataset.
%We check the effect of the number of batches and batch size on the performance of community detection on Pubmed in Figure \ref{fig:batch}.
In Figure \ref{fig:batchesSampled}, we vary the number of batches sampled, where each batch contains 2000 nodes. We can see that the performance improved initially and then saturates when number of batches is 5 or more. Thus, optimal performance on Pubmed for community detection can be achieved by using only \textasciitilde 50\% (or more) of data points to train the self-expressive layer. In Figure \ref{fig:batchSize}, we change the batch size, keeping number of batches as 6. More is the batch size, more computation resource and time needed. We observe that SEComm is able to achieve reasonably good performance when the batch size is 2000 or more.

\subsection{Model Ablation Study}\label{sec:ablation}
In this section, we show the usefulness of different components of SEComm. In particular, we check the community detection performance in the following scenarios. 

\textbf{SEComm-GNN} We run k-means on the node embeddings produced by the self-supervised GNN used in SEComm, without running the other modules of SEComm.

\textbf{SEComm-Spectral} We run spectral clustering on the complete similarity matrix $S$ to find node clusters. However, $S$ can be computed only for smaller graphs and hence this experiment cannot be performed on larger datasets like Pubmed and Physics.

\textbf{SEComm-Embeddings} We run k-means on the node embeddings generated after the complete training of SEComm (including the self-expressive and clustering modules)

We compare the results of the above with the community detection output of the complete model of SEComm in Table \ref{tab:abalResults}. Again we use three metrics clustering accuracy, NMI and F1 score to evaluate the quality of community detection. Interestingly, we do not see any clear winners between the three model variants. But it is clear from the reported performance numbers that the complete model of SEComm outperforms its variants (except Citeseer-F1\%).

\section{Discussion and Conclusion}\label{sec:con}
In this work, we have proposed a novel graph neural network that can directly be used for node community detection in a graph. We use the principle of self-expressiveness to derive a set of soft node-pair constraints to regularize the formation of the communities. To the best of our understanding, this is the first work to integrate a self-expressive layer into a self-supervised GNN. Our approach is highly scalable, without compromising the performance of community detection. SEComm is able to achieve state-of-the-art performance on all the datasets that we used for community detection.

Due to the use of graph neural network to directly generate community memberships of nodes, SEComm can work in an inductive setup. It would be interesting to analyze the performance of SEComm on newly added nodes or even new graphs without retraining.

% \begin{contributions} % will be removed in pdf for initial submission,
%                       % so you can already fill it to test with the
%                       % ‘accepted’ class option
%     Briefly list author contributions.
%     This is a nice way of making clear who did what and to give proper credit.

%     H.~Q.~Bovik conceived the idea and wrote the paper.
%     Coauthor One created the code.
%     Coauthor Two created the figures.
% \end{contributions}

% \begin{contributions}
% Both the authors contributed equally to this research.
% \end{contributions}

\begin{acknowledgements}
We want to thank Prof. M. Narasimha Murty from CSA, IISc for his feedback on this work.
\end{acknowledgements}

\balance
% \bibliography{uai2021-template}
\bibliography{bandyopadhyay_434}

% \appendix
% % NOTE: necessary when ptmx or no mathfont class option is given
% \providecommand{\upGamma}{\Gamma}
% \providecommand{\uppi}{\pi}
% \section{Math font exposition}
% How math looks in equations is important:
% \begin{equation*}
%   F_{\alpha,\beta}^\eta(z) = \upGamma(\tfrac{3}{2}) \prod_{\ell=1}^\infty\eta \frac{z^\ell}{\ell} + \frac{1}{2\uppi}\int_{-\infty}^z\alpha \sum_{k=1}^\infty x^{\beta k}\mathrm{d}x.
% \end{equation*}
% However, one should not ignore how well math mixes with text:
% The frobble function \(f\) transforms zabbies \(z\) into yannies \(y\).
% It is a polynomial \(f(z)=\alpha z + \beta z^2\), where \(-n<\alpha<\beta/n\leq\gamma\), with \(\gamma\) a positive real number.

\end{document}